\documentclass{aastex61}

\accepted{May 10, 2018}

\shorttitle{Coronal Jets Extending to High Altitudes}
\shortauthors{Y. Hanaoka et al.}

\begin{document}

\title{Solar Coronal Jets Extending to High Altitudes Observed During the 2017 August 21 Total Eclipse}

\correspondingauthor{Yoichiro Hanaoka}
\email{hanaoka@solar.mtk.nao.ac.jp}

\author{Yoichiro Hanaoka}
\affil{National Astronomical Observatory of Japan \\
2-21-1 Osawa, Mitaka, Tokyo, 181-8588, Japan}
\affil{Solar Eclipse Digital Imaging and Processing Network \\
Odawara, Kanagawa, 250-0051, Japan}

\author{Ryuichi Hasuo}
\affil{National Astronomical Observatory of Japan \\
2-21-1 Osawa, Mitaka, Tokyo, 181-8588, Japan}
\affil{Solar Eclipse Digital Imaging and Processing Network \\
Odawara, Kanagawa, 250-0051, Japan}

\author{Tsukasa Hirose}
\affil{Solar Eclipse Digital Imaging and Processing Network \\
Odawara, Kanagawa, 250-0051, Japan}

\author{Akiko C. Ikeda}
\affil{Solar Eclipse Digital Imaging and Processing Network \\
Odawara, Kanagawa, 250-0051, Japan}

\author{Tsutomu Ishibashi}
\affil{Solar Eclipse Digital Imaging and Processing Network \\
Odawara, Kanagawa, 250-0051, Japan}

\author{Norihiro Manago}
\affil{Solar Eclipse Digital Imaging and Processing Network \\
Odawara, Kanagawa, 250-0051, Japan}

\author{Yukio Masuda}
\affil{Solar Eclipse Digital Imaging and Processing Network \\
Odawara, Kanagawa, 250-0051, Japan}

\author{Sakuhiro Morita}
\affil{Solar Eclipse Digital Imaging and Processing Network \\
Odawara, Kanagawa, 250-0051, Japan}
\affil{NPO Kwasan Astro Network \\
17-1 Ohmine-cho Kitakazan, Yamashina-ku, Kyoto 607-8471, Japan}

\author{Jun Nakazawa}
\affil{Solar Eclipse Digital Imaging and Processing Network \\
Odawara, Kanagawa, 250-0051, Japan}

\author{Osamu Ohgoe}
\affil{National Astronomical Observatory of Japan \\
2-21-1 Osawa, Mitaka, Tokyo, 181-8588, Japan}
\affil{Solar Eclipse Digital Imaging and Processing Network \\
Odawara, Kanagawa, 250-0051, Japan}

\author{Yoshiaki Sakai}
\affil{Solar Eclipse Digital Imaging and Processing Network \\
Odawara, Kanagawa, 250-0051, Japan}
\affil{Chiba Prefectural Tsurumaisakuragaoka High School\\
Ichihara, Chiba, 290-0512, Japan}

\author{Kazuhiro Sasaki}
\affil{Solar Eclipse Digital Imaging and Processing Network \\
Odawara, Kanagawa, 250-0051, Japan}

\author{Koichi Takahashi}
\affil{NPO Kwasan Astro Network \\
17-1 Ohmine-cho Kitakazan, Yamashina-ku, Kyoto 607-8471, Japan}

\author{Toshiyuki Toi}
\affil{Solar Eclipse Digital Imaging and Processing Network \\
Odawara, Kanagawa, 250-0051, Japan}

\begin{abstract}
Coronal jets, which extend from the solar surface to beyond 2 $R_\odot$, were observed in the polar coronal hole regions during the total solar eclipse on 2017 August 21.  In a time-series of white-light images of the corona spanning 70 minutes taken with our multi-site observations of this eclipse, six jets were found as narrow structures upwardly ejected with the apparent speed of about 450 km s$^{-1}$ in polar plumes.  On the other hand, extreme-ultraviolet (EUV) images taken with the Atmospheric Image Assembly of the Solar Dynamics Observatory show that all of the eclipse jets were preceded by EUV jets.  Conversely, all the EUV jets whose brightness is comparable to ordinary soft X-ray jets and which occurred in the polar regions near the eclipse period were observed as eclipse jets.  These results suggest that ordinary polar jets generally reach high altitudes and escape from the Sun as part of the solar wind. \end{abstract}


\section{Introduction} \label{sec:intro}

Solar coronal jets were first extensively studied using soft X-ray images \citep{1992PASJ...44L.173S, 1996PASJ...48..123S}.  Many studies including some statistical analyses have been carried out using soft X-ray and extreme-ultraviolet (EUV) observations, and now they are understood as common phenomena in the low corona \citep[see e.g.,][]{2016SSRv..201....1R}.  On the other hand, particularly for the jets occurring in polar coronal holes, namely in the open field regions, it is interesting if they escape from the Sun as part of the fast solar wind.  However, it is difficult to determine how far the jets extend from soft X-ray and EUV observations alone.  \citet{1998ApJ...508..899W} found that some jets in the polar regions observed with the Extreme-ultraviolet Imaging Telescope \citep[EIT;][]{1995SoPh..162..291D} on board the Solar and Heliospheric Observatory \citep[SOHO;][]{1995SoPh..162....1D} were also observed in the white-light images beyond 2 $R_\odot$ taken with the Large Angle Spectrometric Coronagraph \citep[LASCO;][]{1995SoPh..162..357B} on board the SOHO.  They found only 27 jets observed with both the EIT and the LASCO in eleven months.  \citet{2009SoPh..259...87N} found that some polar jets were observed both with the Sun Earth Connection Coronal and Heliospheric Investigation (SECCHI)-COR1 coronagraph \citep{2008SSRv..136...67H} and with the SECCHI-Extreme UltraViolet Imager \citep[EUVI;][]{2004SPIE.5171..111W} of the Solar TErrestrial RElations Observatory \citep[STEREO;][]{2008SSRv..136....5K}, and \citet{2015ApJ...806...11M} studied 14 jets in polar coronal holes observed with the LASCO C2 coronagraph, which had the EUV counterpart observed with the Atmospheric Imaging Assembly \citep[AIA;][]{2012SoPh..275...17L} of the Solar Dynamics Observatory \citep[SDO;][]{2012SoPh..275....3P}. Therefore, it is confirmed that at least some jets in the polar regions, probably the highly energetic ones, reach high altitudes in the corona, but it has not been known if the smaller jets do not reach high altitudes or they are too faint at high altitudes to be detected with the spaceborne coronagraphs, as pointed by \citet{2015ApJ...806...11M}. Regarding the relation to the solar wind, \citet{1999ApJ...523..444W} studied the kinematic characteristics of the jets observed with the LASCO C2, and they concluded that the jets seen with the C2 merge into the solar wind and escape from the Sun.  \citet{2010SoPh..264..365P}, who analyzed jets observed with the COR1 and the EUVI, concluded that polar jets contribute to the solar wind.  \citet{2012ApJ...750...50N} argued that the source of the microstream in the solar wind is polar coronal jets.  Therefore, some polar jets, which are probably energetic ones, are considered to escape from the Sun as part of the fast solar wind, but it is not known if ordinary jets escape from the Sun.  

On the other hand, some studies show that jets do not necessarily escape from the Sun but they fall back to the solar surface \citep{2007PASJ...59S.751C, 2014A&A...561A.104C, 2015A&A...579A..96P}.  \citet{2013ApJ...776...16P} showed that ``standard jets'' cannot be seen in the images taken with the COR1 but ``blowout jets'' are sometimes seen.  The classification of standard and blowout jets was introduced by \citet{2010ApJ...720..757M}, and blowout jets have larger width than standard ones, and are more energetic.

One reason that the ultimate height of jets is not very clear is the limitation of the height coverage in the observation of the corona.  As \citet{2016SSRv..201....1R} reviewed, there has only been a small number of observations of jets in high altitudes because of the limited height coverage.  The soft X-ray and EUV instruments mostly cover up to 1.1--1.2 $R_\odot$.  On the other hand, due to the occulting disks of the spaceborne coronagraphs, which are much larger than the solar disk, only the corona beyond 2.0 $R_\odot$ can be observed with the LASCO C2, and that beyond 1.6 $R_\odot$ with the SECCHI-COR1.  Therefore, there is a gap in the height coverage of the corona by the space instruments.  This gap obscures the behavior of EUV and soft X-ray jets beyond 1.1--1.2 $R_\odot$.  Furthermore, it is possible that the height 1.6 -- 2.0 $R_\odot$ is too high for less energetic jets to be detected with spaceborne coronagraphs such as the LASCO and the SECCHI.

At the total solar eclipses, the corona from just above the limb to several solar radii can be observed under the very low sky background level.  Therefore, to observe total solar eclipses is a way to fill the gap mentioned above.  On the other hand, because a total eclipse seen from a certain site lasts only a couple of minutes, it is difficult to track the time variation of the corona.  However, if an eclipse is observed at multiple sites while the eclipse umbra runs along the total eclipse path, the time variation can also be observed.  For example, brightening of a polar ray in the polar plume region was found from six-site observations at the 2006 eclipse \citep{2008ApJ...682..638P}.  At the 2012 eclipse, two-site observations show a development of a coronal mass ejection \citep{2014SoPh..289.2587H}.   Such achievements show the potential of the solar eclipse observations.

On 2017 August 21, a total solar eclipse took place, with its umbra passing across North America for 90 minutes.  We organized a multi-site eclipse observation program and succeeded in taking a time-series of wide dynamic range images of the white-light corona at seven sites. The data covers a time period of about 70 minutes, and such observations enabled us to trace the time variation of the corona from near the solar surface to several solar radii including the gap of the coverage by the spaceborne instruments.  At the eclipse, during the declining phase of the solar cycle 24, the solar activity was already low.  The polar coronal holes were remarkable and the polar plume structures developed well above them.  The data obtained at the eclipse are suitable to investigate unseen behavior of polar jets above the height observed with the EUV and soft X-ray instruments.   In this paper, we present an analysis of the data of polar jets taken at the eclipse in conjunction with the data simultaneously obtained with the spaceborne instruments.  The observations are described in Section 2, and the results of the analysis of the eclipse data and the comparison with the data taken with the spaceborne instruments are presented in Section 3.  Discussion and a summary are given in Section 4.

\section{Observations and Data} \label{sec:obs}

\begin{table}
\caption{Eclipse observations \label{tab:table1}}
\begin{tabular}{ccll}
\hline\hline
Time of  & Telescope aperture & Camera & Observation site \\
maximum eclipse & and focal length & \\
(UT) & (mm) & & \\
\hline
17:18 & D60 f600 & Nikon D810a & Salem, Oregon, USA \\
17:20 & D60 f600 & Canon EOS6D & Madras, Oregon, USA \\
17:26 & D76 f594 & Nikon D500 & Weiser, Idaho, USA \\
17:34 & D60 f600 & Nikon D810a & Rexburg, Idaho, USA \\
17:46 & D54 f300 & Canon EOS6D & Glendo, Wyoming, USA \\
18:09 & D71 f560 & Nikon D810a & Excelsior Springs, Missouri, USA \\
18:28 & D85 f680 & Canon EOS5DmkII & Nashville, Tennessee, USA \\
\hline\hline
\end{tabular}
\end{table}

At the total solar eclipse on 2017 August 21, professionals and amateurs in Japan collaboratively performed a multi-site observation program.  The collaboration to observe the white-light corona at total solar eclipses has lasted nearly ten years, and it has produced some scientific results \citep{2012SoPh..279...75H, 2014SoPh..289.2587H}.  At this eclipse, observers succeeded in obtaining time-series white-light images of the corona at seven sites, with most taken with several different exposure times to achieve a wide dynamic range.  Table 1 summarizes the eclipse observations.  The observations were done with small refractor telescopes and single lens reflex cameras, which covers up to several solar radii of the white-light corona.  The observation at 17:46 UT was done with a single exposure time. Nevertheless we included this one to fill the gap between 17:34 UT and 18:09 UT, because in the eastern (latter) half of the eclipse path the intervals of the observations became long due to rather poor weather conditions.

A set of data taken at an observation site with various exposure times is composited to produce a white-light image with a wide dynamic range.  In the case that the analog-to-digital conversion in a camera shows non-linear behavior, we corrected it for each image before the composite.  The wide dynamic range enables us to study the coronal structure from just above the limb to several solar radii.  The composite images obtained from the observations at the seven sites have different brightness scales, and they have different scattered light levels depending on the sky conditions of the observing sites.  We calibrated the brightness scale and corrected the sky background so that we could find the changes in the corona during the observations on the basis of the time-series images.

The eclipse data were compared to the data obtained with spaceborne instruments.  The AIA of the SDO takes full-Sun images including the low corona up to about 1.2 $R_\odot$ every 12 s with the spatial sampling of $0''.6$ pixel$^{-1}$ using various filters in the UV--EUV wavelengths, which correspond to the temperature range of 0.05--20 MK.  Because the SDO was not eclipsed during our eclipse observations, we were able to check the activity near the solar surface such as jets during the eclipse period using EUV images.  We mainly used reduced size ``synoptic series'' data ($2''.4$ pixel$^{-1}$, every two minutes) for our analysis, and also used full-resolution data as needed.  The soft X-ray observation has been a standard method to detect coronal jets, and therefore, we checked the data taken with the X-Ray Telescope \citep[XRT;][]{2007SoPh..243...63G} on board {\it Hinode} \citep{2007SoPh..243....3K} as well.  The XRT has a field of view covering the full Sun with the spatial sampling of $1''.0286$.  However, until 16:38 UT the XRT were taking partial images around the north pole with the full resolution ($1''.0286$ pixel$^{-1}$), then it took the full-Sun images with a reduced resolution ($2''.0572$ pixel$^{-1}$) at 16:43 UT and from 17:10 UT on.  Therefore, there are significant gaps in the XRT observation just before the eclipse.

Jets are also checked in the images taken with the LASCO C2 coronagraph of the SOHO.  The field of view of the LASCO C2 covers up to 6 $R_\odot$ and it shows the white-light corona extending farther than the eclipse image.  However, the occulting disk covers up to 2 $R_\odot$, and about within 2.5 $R_\odot$ it is difficult to see the structure of the corona.

\section{Results} \label{sec:res}

\subsection{Jets found in the eclipse data}

\begin{figure}
\plottwo{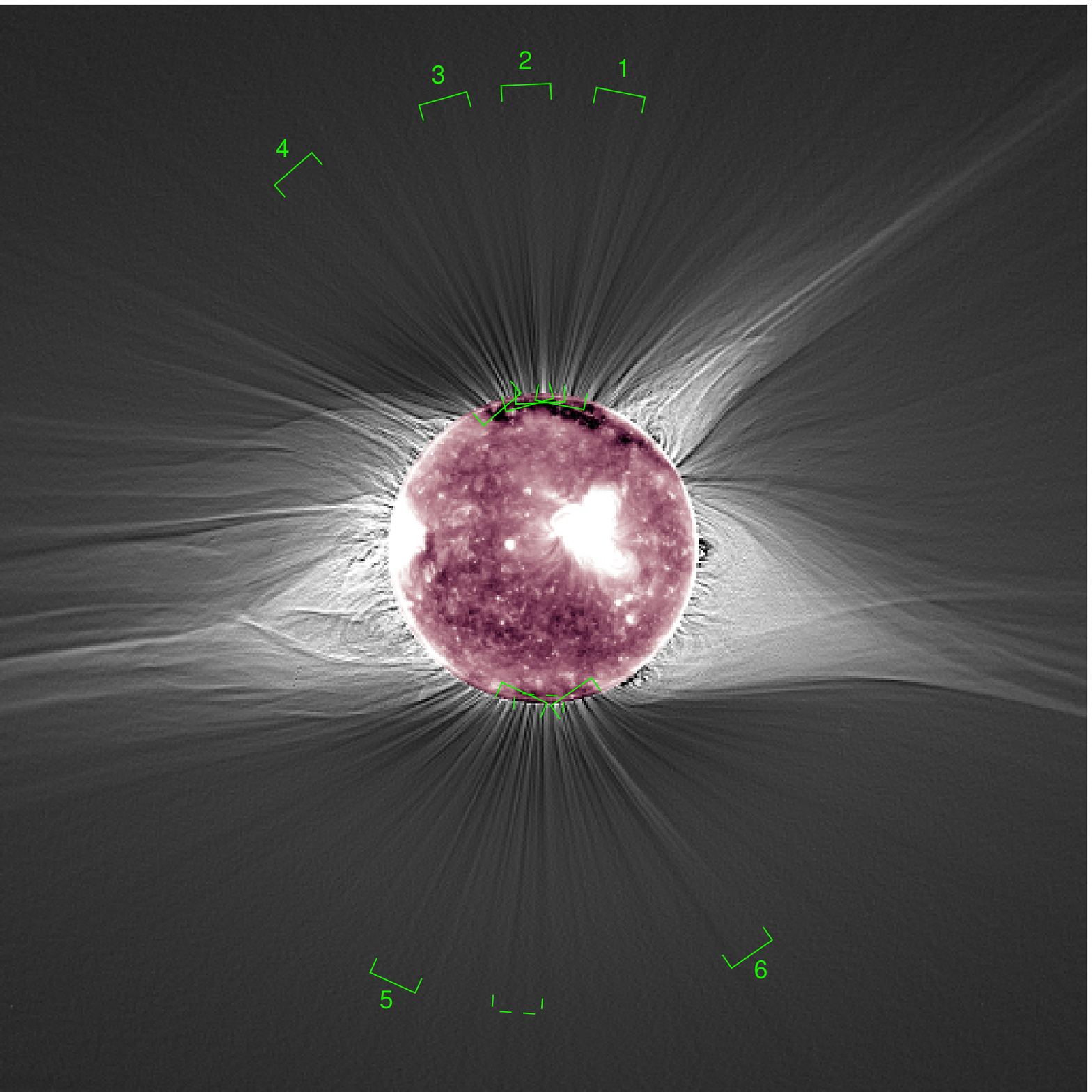}{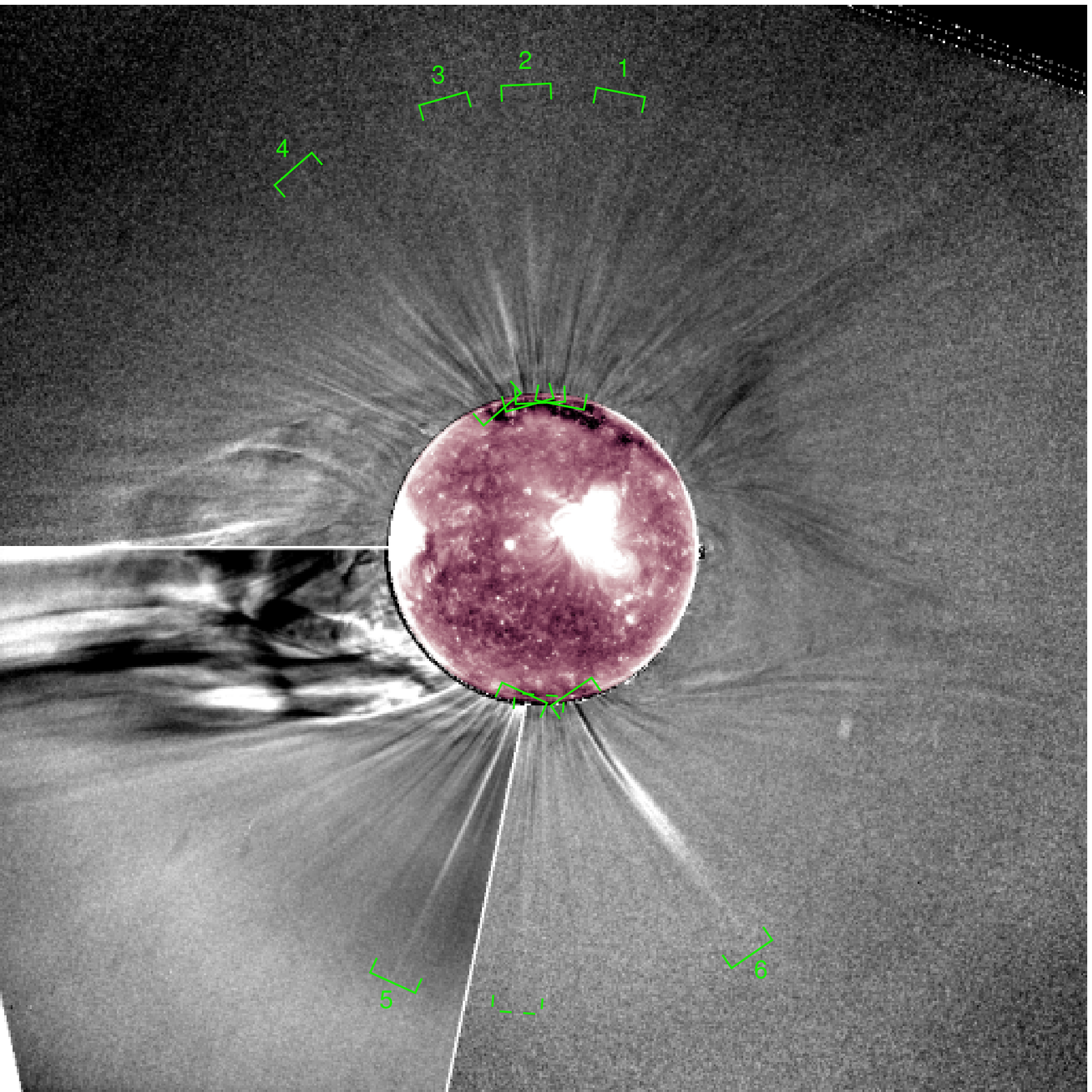}
\caption{
White-light corona and positions of the eclipse jets found in the white-light images taken during the solar eclipse on 2017 August 21.  An EUV image at 211 \AA\ taken with the AIA of the SDO is shown instead of the Moon.  The solar north is to the top.  (left) White-light image, where coronal fine structures are enhanced.  Six strips, which include the positions of the jets, are also shown.  The numbers correspond to the jet numbers explained in Section 3.1.  In addition, a strip marked by dashed lines shows the position of a possible eclipse jet described in Section 3.2.  (right) Difference between the white-light images taken at 17:18 UT and 17:34 UT. Brightenings and darkenings caused by various kinds of active phenomena including the jets can be seen.  The lower-left section is replaced with the difference between the images at 17:34 UT and 18:28 UT to show jet 5 (see Section 3.1) clearly.  Strips in the left panel are shown here again.
\label{fig:fig1}}
\end{figure}

\begin{figure}
\plotone{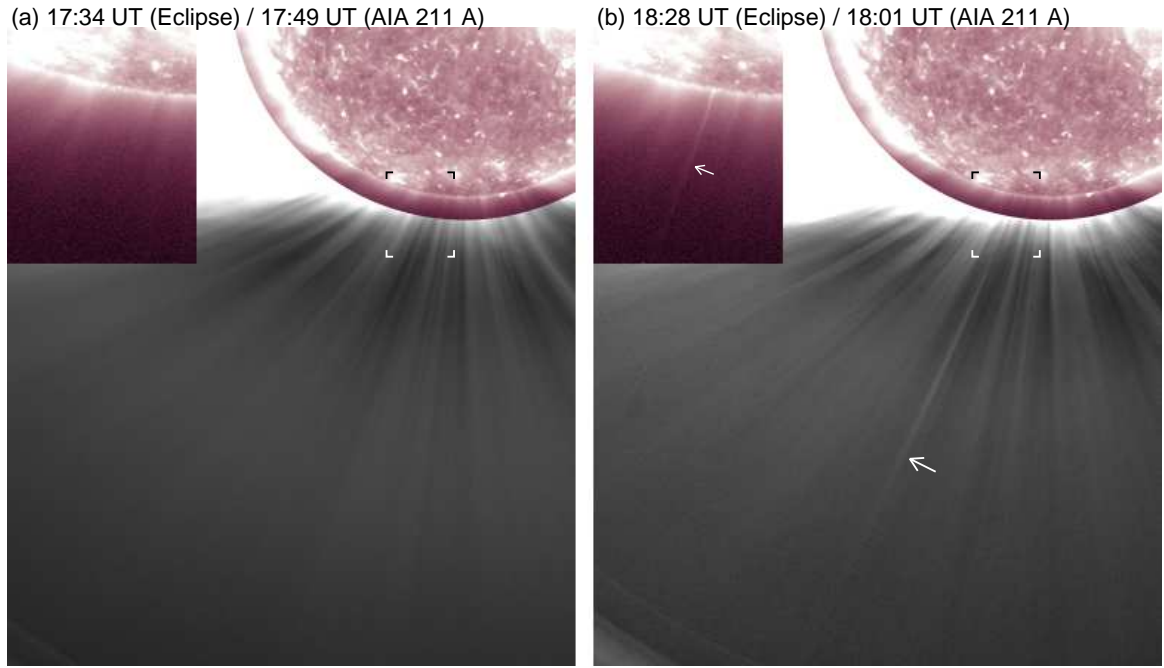}
\caption{
EUV images at 211 \AA\ taken with the AIA of the SDO and eclipse white-light images before and after the occurrence of a jet (jet 5 described in Section 3.1).  (a) EUV (17:49 UT) and white-light (17:34 UT) images before the jet.  A box in the image includes the position of the jet, and an enlargement of the EUV image in this box is shown at the upper-left corner.  (b) EUV (18:01 UT) and white-light (18:28 UT) images including the jet, which is indicated by an arrow.  An enlarged EUV image is also shown at the upper-left corner, and the jet is indicated by an arrow.  The white-light images are processed to suppress the steep radial brightness gradient and to enhance the jet.
\label{fig:fig2}}
\end{figure}

\begin{deluxetable*}{ccccccc}
\tablenum{2}
\tablecaption{Parameters of the jets \label{tab:table2}}
\tablewidth{0pt}
\tablehead{
\colhead{Jet number} & \colhead{Position angle} & \colhead{Start time} & \colhead{Peak time} & \colhead{$v_\mathrm{EUV}$} & \colhead{$v_{\mathrm{EUV}+\mathrm{eclipse}}$} & \colhead{$v_\mathrm{eclipse}$} \\
\nocolhead{} & \colhead{(degree)} & \colhead{(UT)} & \colhead{(UT)} & \colhead{(km s$^{-1}$)} & \colhead{(km s$^{-1}$)} & \colhead{(km s$^{-1}$)} 
}
\startdata
1 & 353 & 16:43 & 16:51 & 340 & 270 & 250 \\
2 & 1 & 17:04 & 17:10 & 360 & 450 & 130 \\
3 & 6 & 17:03 & 17:13 & 290 & 410 & 230 \\
4 & 17 & 16:55 & 17:05 & 250 & 350 & 200 \\
5 & 172 & 17:53 & 18:01 & 660 & 730 & 300 \\
6 & 192 & 16:53 & 17:07 & 290 & 500 & 500 \\
average &  &  &  & 360 & 450 & 270 \\
\enddata
\tablecomments{The start time and the peak time are those of the EUV jets.  The peak time is when a jet is most conspicuously seen. The jet velocities, $v_\mathrm{EUV}$, $v_{\mathrm{EUV}+\mathrm{eclipse}}$, and $v_\mathrm{eclipse}$, correspond to the values estimated only from the EUV images, those from the length of the jets in the eclipse images and the time elapsed from the EUV start time to the eclipse observations, and those only from the eclipse images, respectively.}
\end{deluxetable*}

\begin{figure}

\gridline{\fig{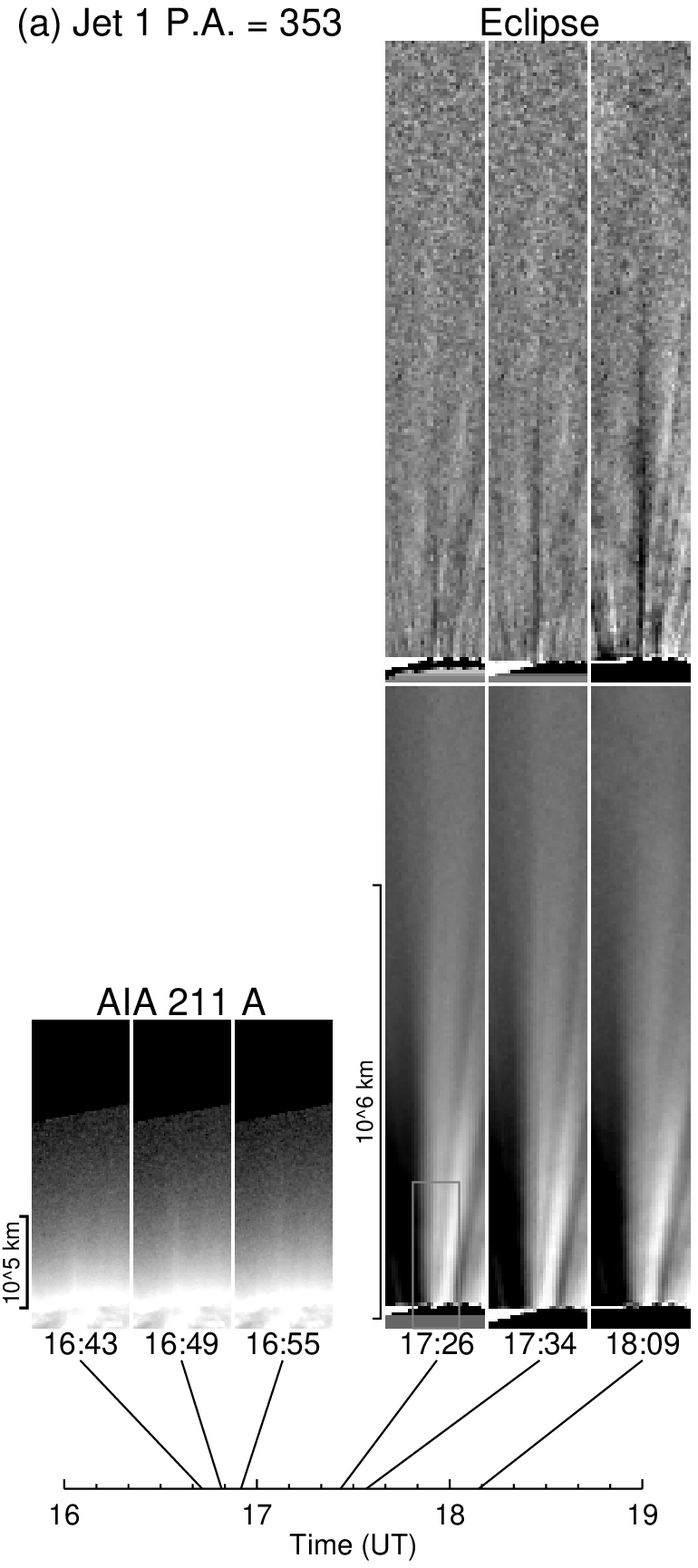}{0.22\textwidth}{}
           \fig{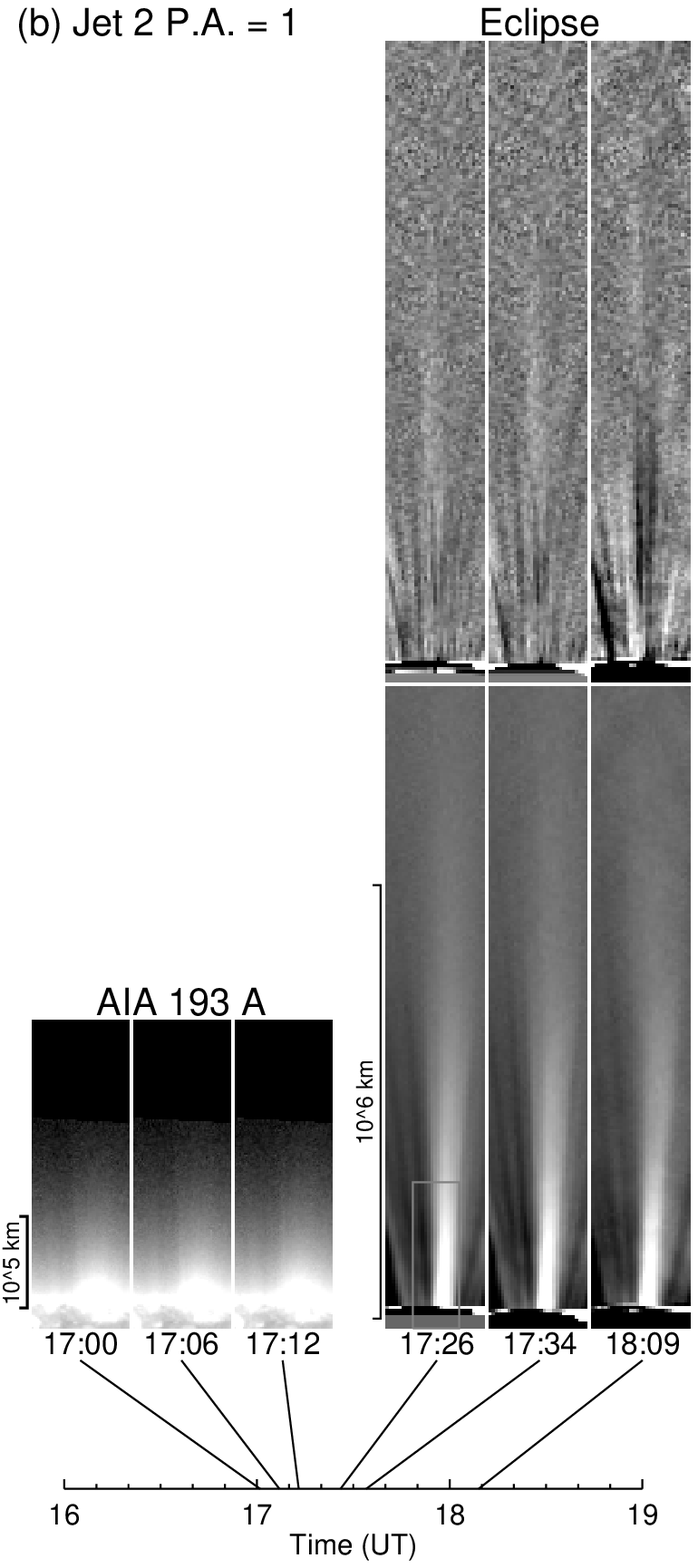}{0.22\textwidth}{}
           \fig{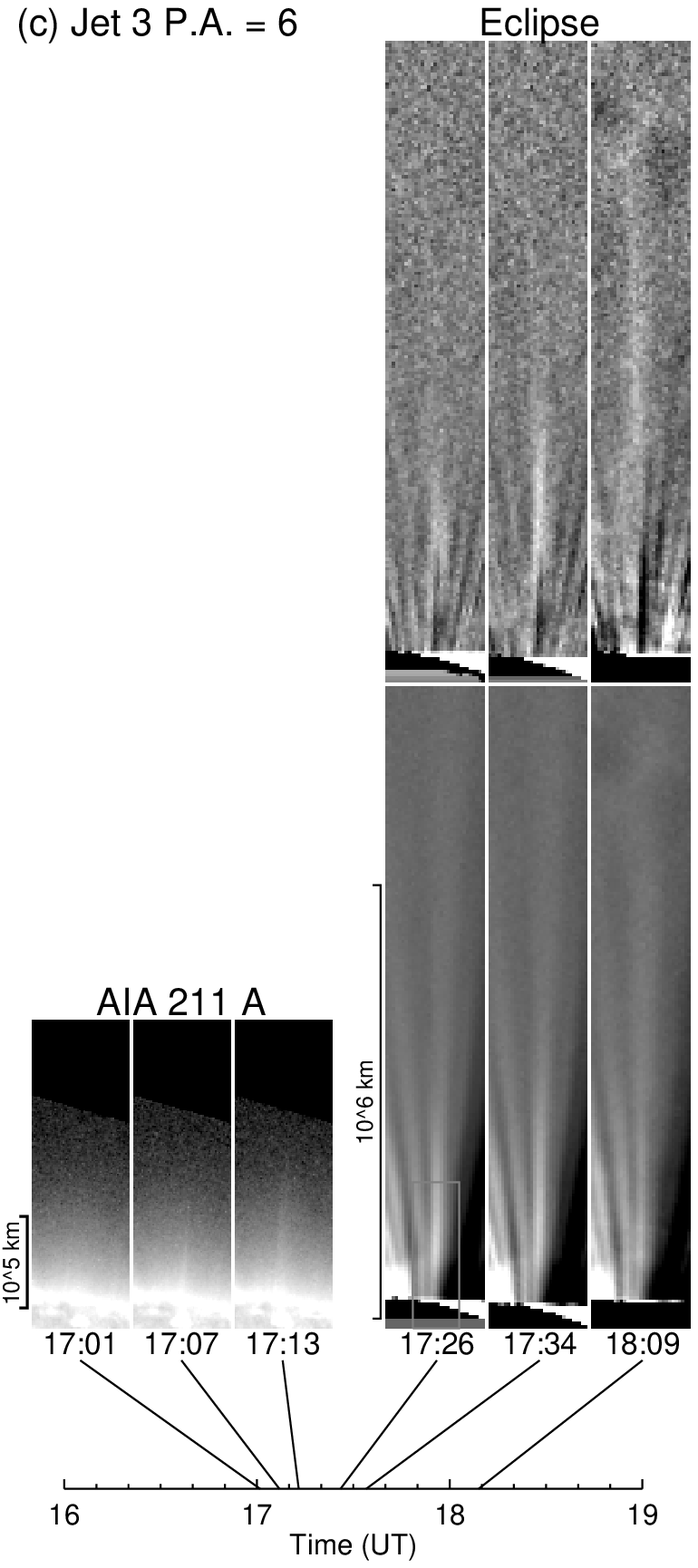}{0.22\textwidth}{}
           \fig{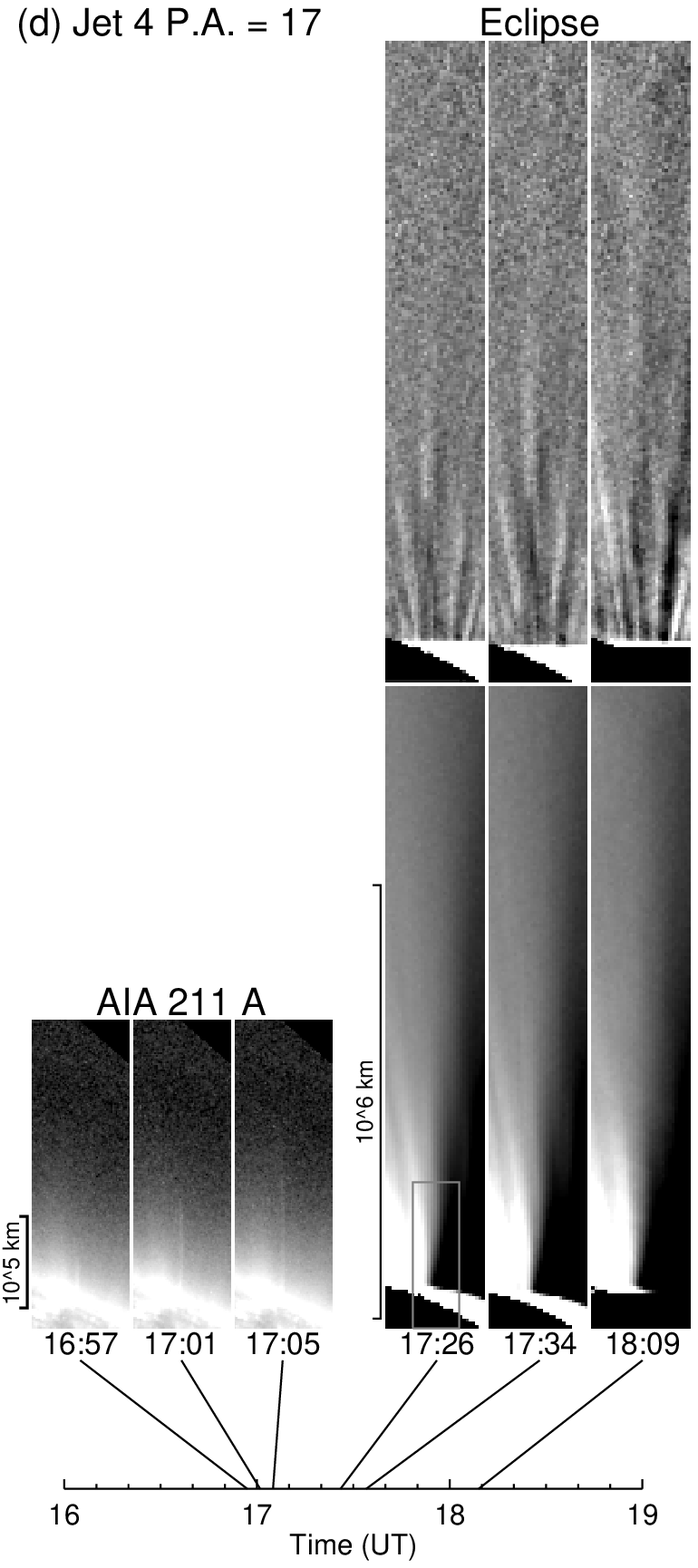}{0.22\textwidth}{}
}
\gridline{\fig{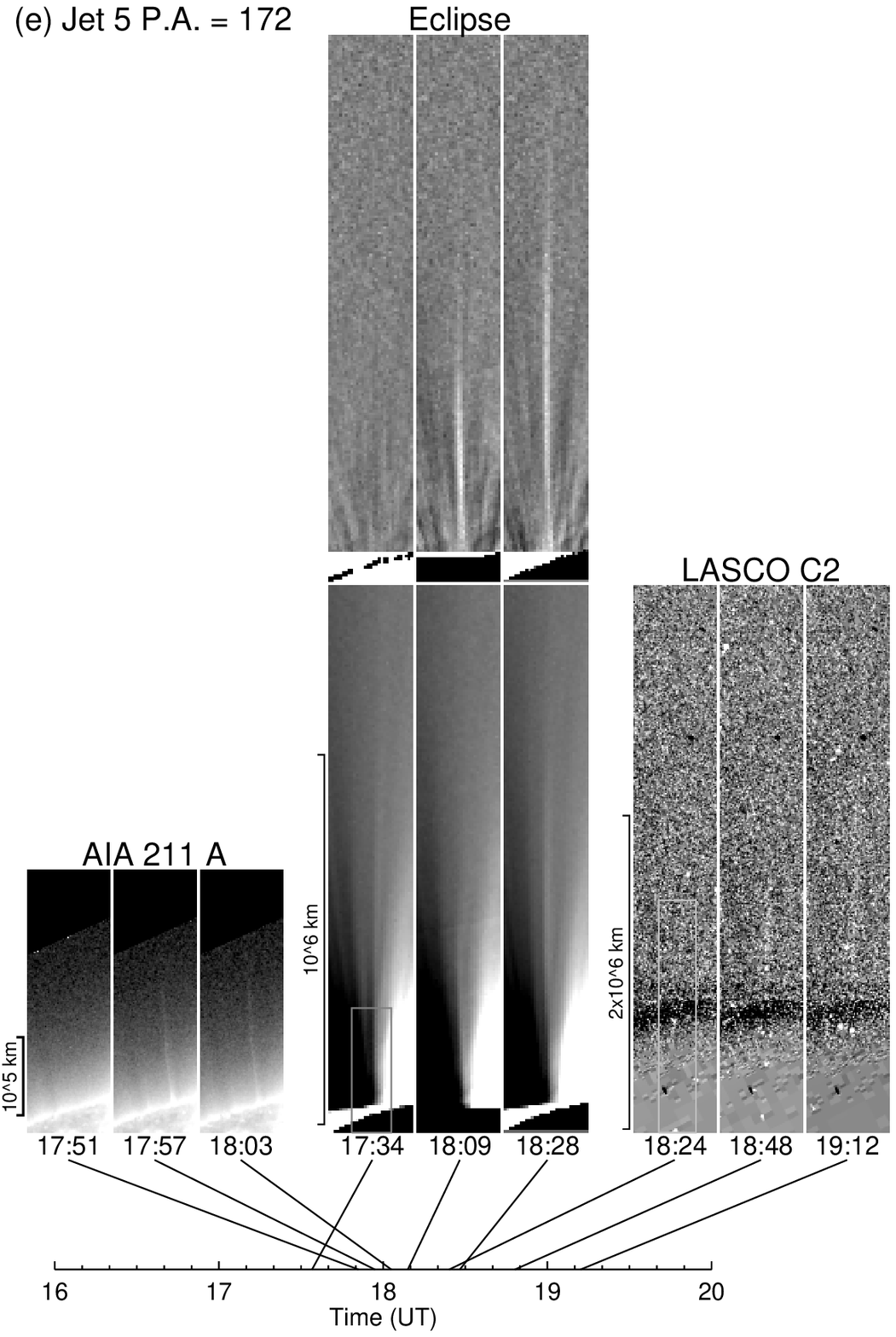}{0.33\textwidth}{}
           \fig{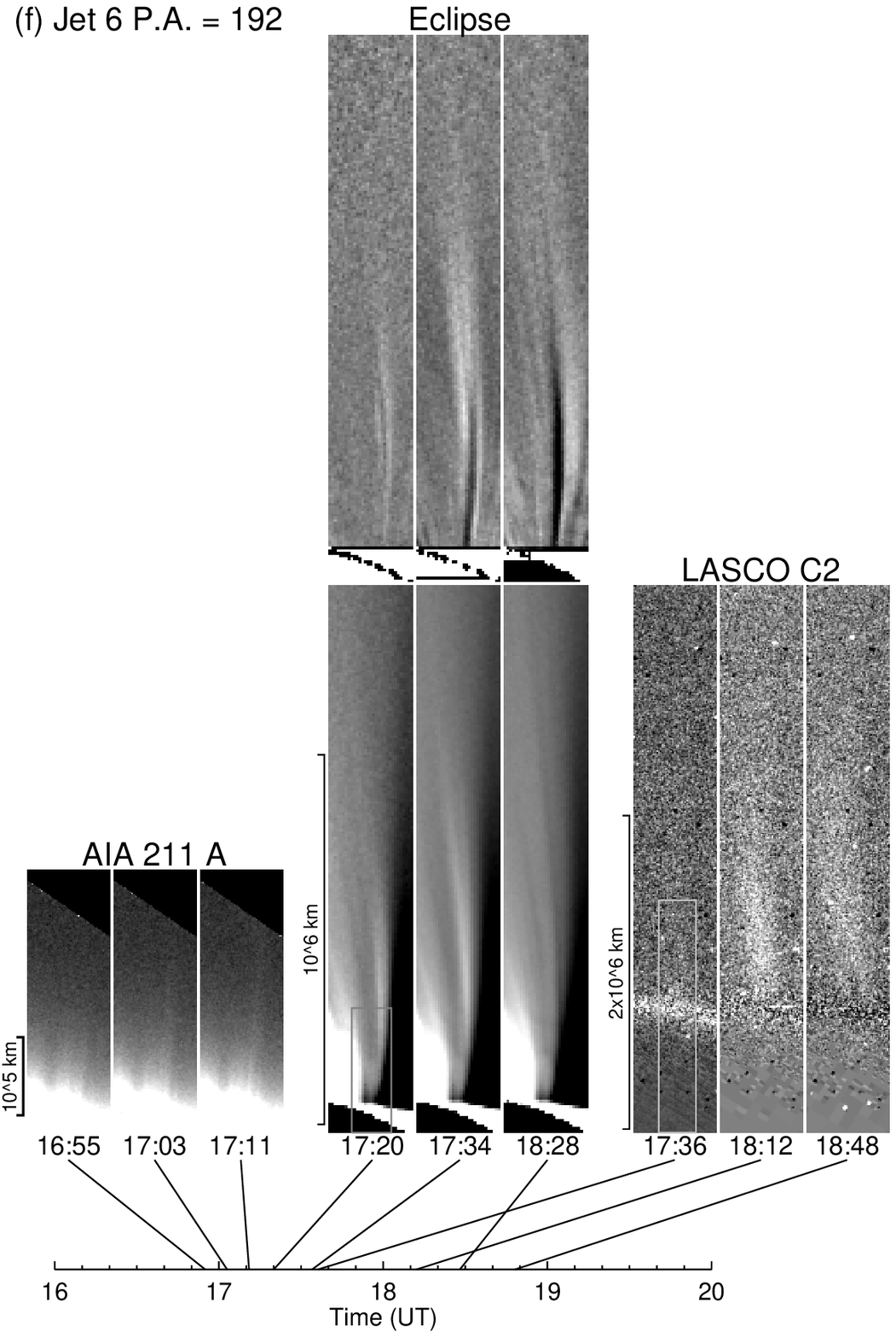}{0.33\textwidth}{}
           \fig{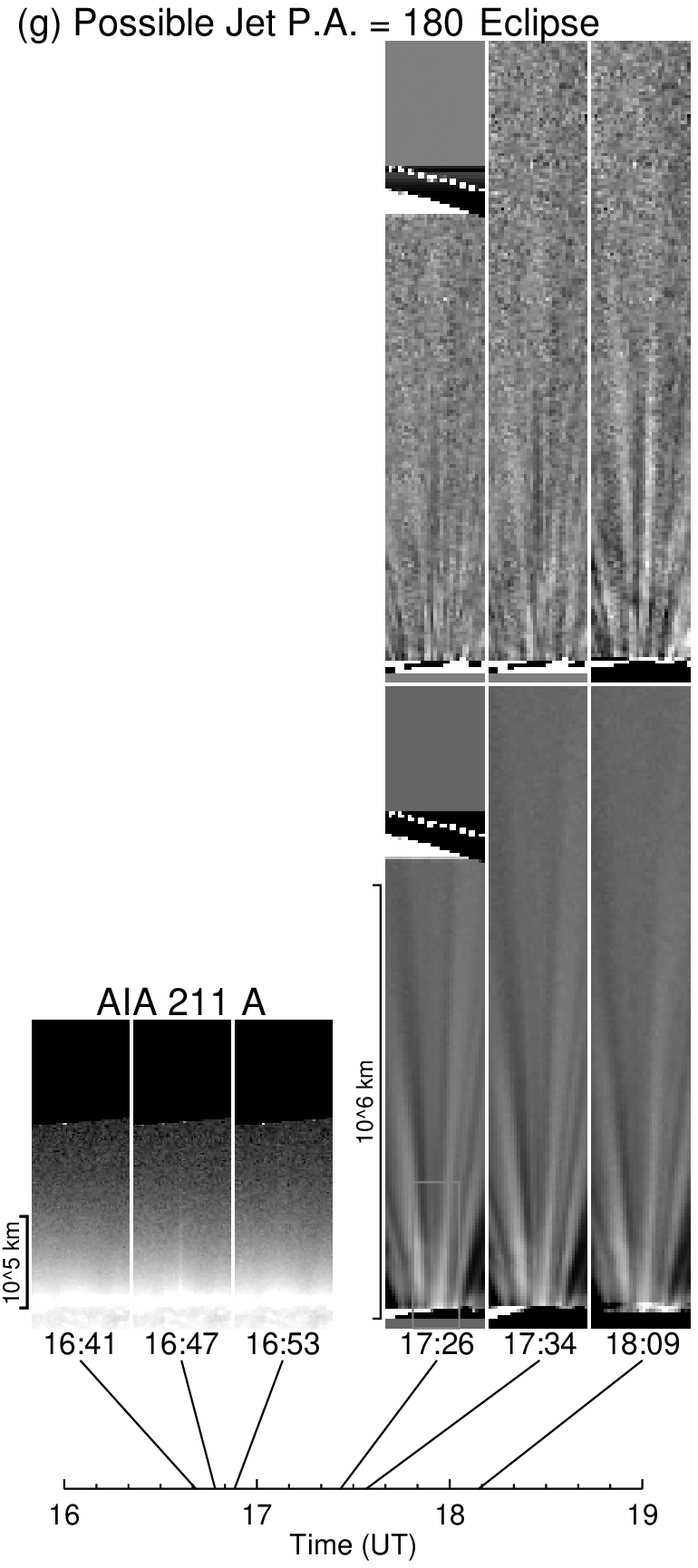}{0.22\textwidth}{}
}
\caption{
Images of the jets taken with the AIA of the SDO, those taken at the eclipse observations, and those taken with the LASCO C2 of the SOHO, which demonstrate the evolution of the jets.  Panels (a)--(f) show jets 1--6 (see Section 3.1), and panel (g) show a possible eclipse jet described in Section 3.2.  The EUV images at 211 \AA\ (193 \AA\ for panel (b)) are shown on the left.  The position of the field of view of the EUV images is delineated by a box in one of the eclipse images.  White-light images taken at the eclipse are shown on the right (center in panels (e) and (f)).  The strips shown in Figure 1 correspond to the field of view of the eclipse images in this figure.  The lower half shows the brightness of the white-light corona (the steep radial brightness gradient is suppressed), and the upper half shows the difference from the reference image taken at 17:18 UT.  The difference was normalized by the brightness of the reference image.  In panels (e) and (f), white-light difference images taken with the LASCO C2 are shown on the right.  The C2 reference image is at 17:48 UT for panel (e) and 17:24 UT for panel (f).  The position of the field of view of the eclipse images is also delineated by a box in one of the LASCO C2 images.  The timeline of the acquisitions of the images is shown at the bottom.
\label{fig:fig3}}
\end{figure}

\begin{figure}
\epsscale{0.5}
\plotone{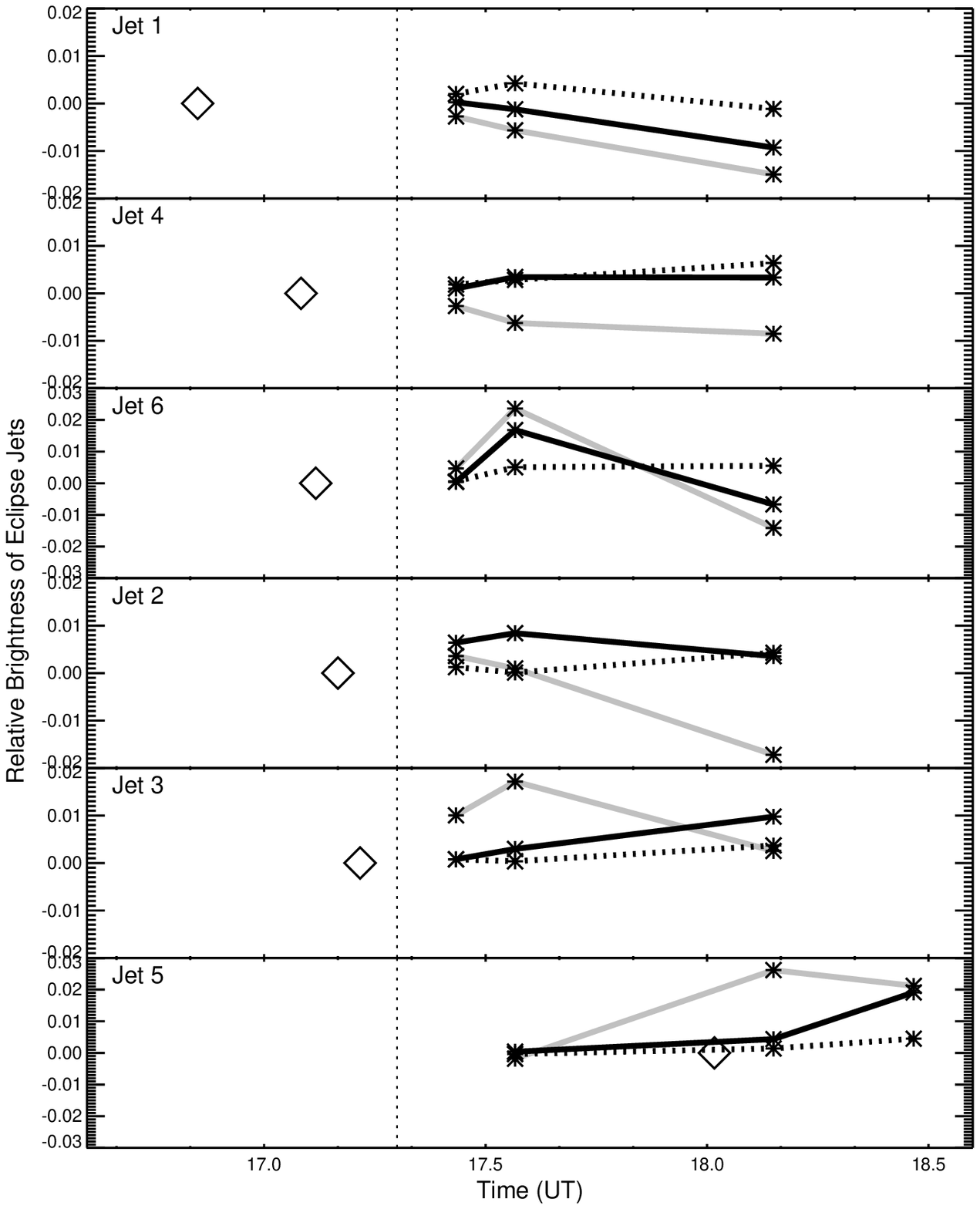}
\caption{Relative brightness changes of the eclipse jets with respect to the reference white-light image taken at 17:18 UT (marked with a vertical dotted line).  The brightness is measured at the axes of the jets at the distances of $3.5\times 10^5$ km (gray lines), $7.0\times 10^5$ km (black solid lines), and $10.5\times 10^5$ km (dotted lines) from the footpoints of the jets.  The brightness values measured at 17:20 UT, 17:34 UT, and 18:09 UT for the jets except jet 5, those at 17:34 UT, 18:09 UT, and 18:28 UT for jet 5, are plotted.  The jets are displayed from the top to the bottom in the order of the peak time of the corresponding EUV jets, which are marked with diamonds.
\label{fig:fig4}}
\end{figure}

The time-series white-light images taken at the eclipse show the changes in the corona over about 70 minutes.  In the ordinary brightness images, only a small number of changes can be recognized, but in the difference images we can find various changes including a small coronal mass ejection above the east limb and brightenings in some streamer structures above active areas.  Polar plumes also show changes, but they are possibly apparent ones caused by the change in the overlaps of the plumes along the line of sight due to the solar rotation, because numerous plumes fill over polar coronal holes.  However, if a change is continually seen over more than three epochs, it can be considered an intrinsic change in the plumes.  Among such changes, we found six upward-moving narrow streams.  Their positions are shown in Figure 1.  In particular, the difference image in the right panel shows various changes in polar plumes, and six of them were identified as ascending structures.  In some cases ascending brightenings are found, while in others, ascending darkenings are found.  The ascending speed is the order of several hundred kilometers per second.  The ascending of collimated narrow streams with this speed is similar to coronal white-light jets, which have been extensively studied by \citet{1998ApJ...508..899W}, \citet{2009SoPh..259...87N}, and \citet{2010SoPh..264..365P}.  Therefore, these are considered to be jets seen on the eclipse data, or eclipse jets.  Ascending darkenings are considered as the dimming after the brightening due to jets, which also proceeds from the footpoint of the jet to a high altitude.

Then we checked EUV images taken with the AIA to find events in the low corona related to these eclipse jets.  We found the corresponding EUV jets for all the eclipse jets on the images taken with the 211 \AA\ filter (the characteristic temperature log $T$ is 6.3) and/or the 193 \AA\ filter (log $T$ is 6.1 and 7.3).  Figure 2 shows a representative example of a jet observed both in the EUV and the eclipse.  An EUV jet appearing on 18:01 UT extended high and it was seen in the eclipse image at 18:28 UT as a high altitude white-light jet reaching nearly 3 $R_\odot$.  This is one of the first examples of jets imaged from the solar surface to a high altitude contiguously, while \citet{2008ApJ...682..638P} found brightening in a polar ray propagating upward with a much slower velocity of 65 km s$^{-1}$ at the 2006 eclipse.   All the six jets seen in the eclipse images and the corresponding EUV jets are shown in panels (a)--(f) of Figure 3.  In Figure 3, the jets in Figure 1 are displayed in the counterclockwise order from the northern hemisphere to the southern hemisphere.  In addition, we checked LASCO C2 images, and we found two corresponding jets.  They are also shown in Figure 3.  The details of these jets are described below.  The various parameters of the jets are compiled in Table 2.

(1) Jet 1 at PA $= 353^\circ$

Figure 3(a) shows a jet whose footpoint is located at position angle (measured counterclockwise from the north pole) 353$^\circ$ in the northern polar region. 
This jet first appeared at 16:43 UT, and most conspicuously seen at 16:51 UT (``peak time'' in Table 2) in the EUV images.   Because the EUV jets have different durations, we visually determined the ``peak time'' of the EUV jets, when the jets are most conspicuously seen, to designate the occurrence time of jets tentatively.  Even though the ``peak time'' is not necessarily corresponds to the peak of the brightness or the length in the EUV, but it represents the time when the EUV jets were observed.  On the left of Figure 3(a), three AIA 211 \AA\ images of an equal time interval (not necessary includes of the peak of the jet), which demonstrate the evolution of the EUV jet, are displayed.  The majority of the EUV jets are most remarkably seen in the 211\AA\ images.
Three sets of eclipse images, which also show the evolution of the eclipse jets, are displayed on the right.  The field of view of the eclipse jet images corresponds to the strip ``1'' marked in Figure 1.  The lower half is the brightness images, and the upper half is the difference images with respect to the reference image taken at 17:18 UT (this is the common reference image for all the jets). The difference images are normalized by the reference image, and therefore, they show the ratio of the brightness difference with respect to the brightness of the reference image.  The timeline of the EUV and eclipse observations is shown below.  In the eclipse images, the jet is not clear in the brightness images, but the difference images show an upward-propagating dark structure.  The reference image of 17:18 UT was taken 27 minutes after the peak of the EUV jet, 16:51 UT.  In the case of the jet shown in Figure 2, the eclipse jet has already reached nearly 3 $R_\odot$ at 18:28 UT, 27 minutes after the observation time of the EUV jet.  Therefore, it is presumed that the jet in Figure 3(a) was already well elongated at 17:18 UT to a high altitude, and the process of the darkening of the jet subsequently to the brightening is seen in the difference images.  The leading edge of the EUV jet, which was defined by visual inspection, extends with the velocity of about 340 km s$^{-1}$.  The length of the darkening in the eclipse image and the time elapsed from the start of the EUV jet give the ascending velocity of about 270 km s$^{-1}$.  However, because the leading edge of the brightening of the white-light jet is probably higher than the observed leading edge of the darkening, this velocity is the lower limit.  The extension of the length of the jet on the eclipse images gives the velocity of about 250 km s$^{-1}$.  In Table 2, the velocities estimated for each jet and the average values of the six jets are presented. 

(2) Jet 2 at PA $= 1^\circ$

Figure 3(b) shows a jet found at position angle of $1^\circ$ in the northern polar region.  On the left, 193 \AA\ images taken with the AIA are shown, because the jet is not very clear in the 211 \AA\ images.  The difference images of the eclipse white-light corona show an ascending brightening and a subsequent darkening.  The reason why the brightening phase of the jet was observed in the white-light images is that the peak time of the EUV jet, 17:10 UT, is well later than that of jet 1.

(3) Jet 3 at PA $= 6^\circ$

Figure 3(c) shows a jet found at position angle of $6^\circ$ in the northern polar region.  On the left, 211 \AA\ images are shown.  In this jet, the difference images again show an ascending brightening and a subsequent darkening, and this is consistent with the fact that the peak time of the EUV jet, 17:13 UT, is close to that of jet 2.

(4) Jet 4 at PA $= 17^\circ$

Figure 3(d) shows a jet found at position angle of $17^\circ$ in the northern polar region.  On the left, 211 \AA\ images are shown.  In this jet, the difference images again show an ascending brightening and a subsequent darkening.  The peak time of the EUV jet, 17:05 UT, is close to those of jets 2 and 3.

(5) Jet 5 at PA $= 172^\circ$

Figure 3(e) shows a jet found at position angle of $172^\circ$ in the southern polar region, which was already presented in Figure 2.  On the left, 211 \AA\ images are shown.  The peak time of the EUV jet is as late as 18:01 UT, and therefore, we were able to catch this jet from before the start in the eclipse observations. In the difference image, the brightening is clearly seen.  On the right, the difference images taken with the LASCO C2, whose reference is the image at 17:48 UT, are shown.  In the image at 18:24 UT, close to the time of the last eclipse image, there is no brightening, but later, we can marginally find an extending faint bright structure in the C2 images.  The height of the C2 jet and the time elapsed from the start of the EUV jet give the ascending velocity of about 480 km s$^{-1}$, which is somewhat smaller than the velocity estimated from the height of the eclipse jet and the elapsed time, 730 km s$^{-1}$.

(6) Jet 6 at PA $= 192^\circ$

Figure 3(f) shows a jet found at position angle of $192^\circ$ in the southern polar region.  On the left, 211 \AA\ images are shown.  It looks like a faint jet in the 211 \AA\ image, but unlike the fine thread-like jets above, the width of this jet is as wide as $3\times 10^4$ km, and integrated brightness is much larger than that of the above EUV jets.  The peak time of the EUV jet is 17:07 UT, which is similar to those of jet 2--4.  However, this jet is much brighter than jets 2--4 in the eclipse images, and therefore, we can track the evolution of the eclipse jet until the last image taken at 18:28, while jets 2--4 disappeared in the latter half of the eclipse observations.  The white-light images taken with the LASCO C2 also show a brightening.  On the right, the C2 difference images, whose reference is the image at 17:24 UT, are shown.  At 18:12 UT, about one hour later than the peak time of the EUV jet, we can find an extending faint bright structure in the C2 field of view.  The length of the C2 jet and the time elapsed from the start of the EUV jet give the ascending velocity of about 450 km s$^{-1}$.   This jet is presumed to be similar to those studied by Wang et al. (1998), who studied the jets observed both with the EIT and LASCO.

To get a general view of the above results, we plotted the relative brightness changes of the eclipse jets at the distances of $3.5\times 10^5$ km, $7.0\times 10^5$ km, and $10.5\times 10^5$ km from the footpoints of jets in Figure 4.  These distances correspond to the heights of 1.5, 2.0 and 2.5 $R_\odot$ (approximately; note that the jets do not elongate exactly radially).  The jets are displayed in the order of the peak time of the corresponding EUV jets (marked with diamonds).  From this figure, we can find a tendency of the brightness changes of the eclipse jets.  At the distance of 1.5 $R_\odot$ (gray lines), all the jets show the brightness decrease at least in the second interval of the plots.  At the height of 2.0 $R_\odot$ (black solid lines), while jet 1 shows the monotonic decrease of the brightness, jets 4, 6, and 2 show the brightening in the first interval of the plots, and furthermore, jets 3 and 5 show the monotonic increase of the brightness.  At the height of 2.5 $R_\odot$ (black dotted lines), the brightness increase in the second interval of the plots can be found in all the jets except jet 1.  As a whole, the jets first appear (and disappear) in the EUV, then the brightening and the subsequent darkening proceed from lower to higher positions in the eclipse jets.  In some cases they appear in the LASCO C2 field of view later on as shown in Figure 3.  These results signify the one-way upward motion of the jets, and there is no observational evidence for the fallback of the jet material to the solar surface.  Therefore, it is presumed that the observed jets drain away beyond the limit of the observations and eventually escape from the Sun as part of the fast solar wind.

\subsection{EUV and soft X-ray jets corresponding to and not corresponding to the eclipse jets}

\begin{figure}
\plotone{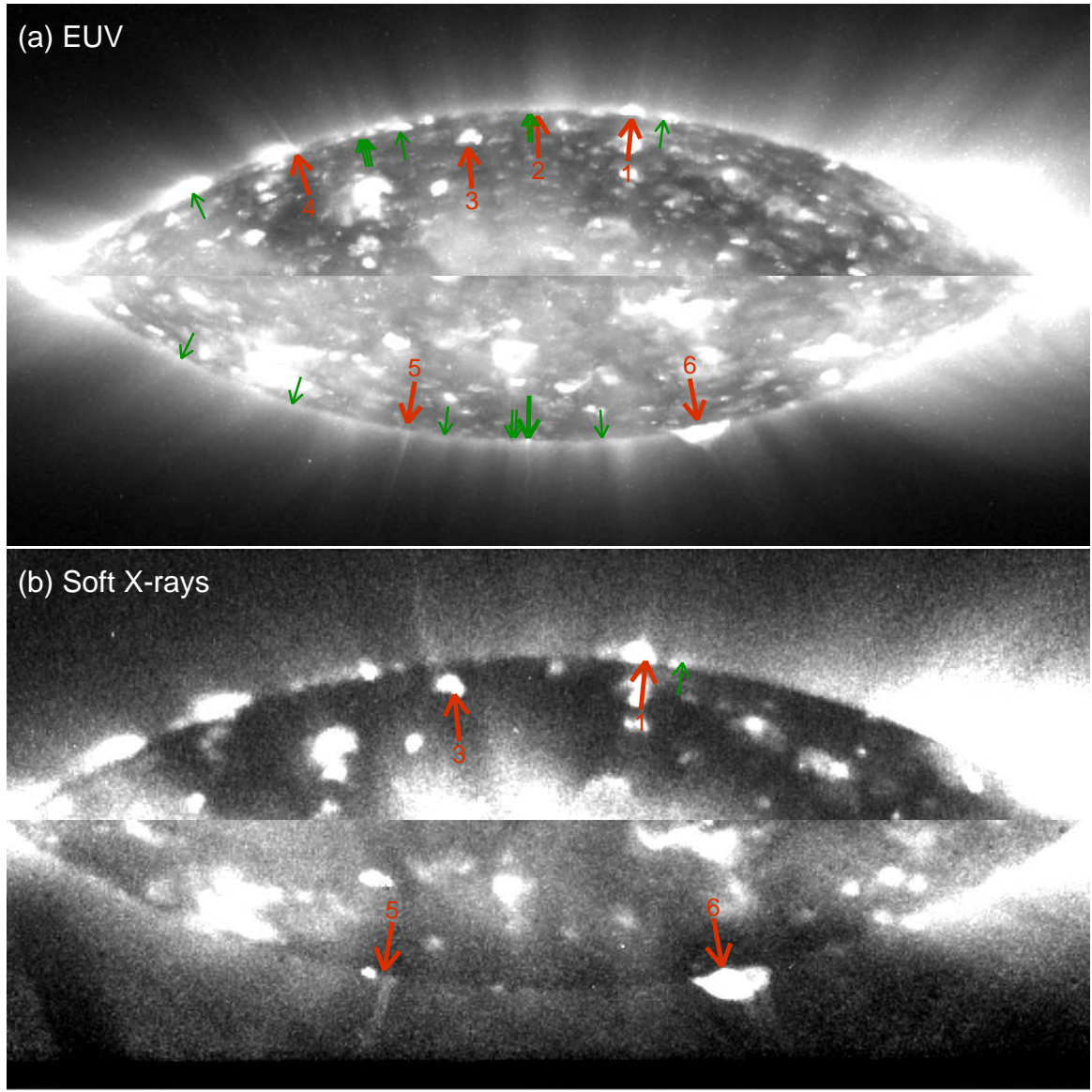}
\caption{Positions of the footpoints of the EUV and soft X-ray jets found in the northern and southern polar regions during 16:00 -- 19:00 UT on 2017 August 21.  (a) The footpoints of the EUV jets plotted on a 211 \AA\ image of the polar region corona taken with the AIA of the SDO.  This image was synthesized from the 211\AA\ images taken during 16:00 -- 19:00 UT to enhance the jets.  Red arrows correspond to ones accompanied by the eclipse jets (the jet numbers are also shown), and green arrows correspond to ones not accompanied by the eclipse jets.  The big green arrow near the south pole shows the footpoint of the possible eclipse jet described in Section 3.2.  (b) The footpoints of the soft X-ray jets plotted on a soft X-ray image of the polar region corona taken with the XRT of Hinode.  This image was synthesized from the images including the jets to enhance the jets.  Red arrows correspond to ones accompanied by the eclipse jets (the jet numbers are also shown), and a green arrow corresponds to one not accompanied by the eclipse jets.
\label{fig:fig5}}
\end{figure}

\begin{figure}
\plotone{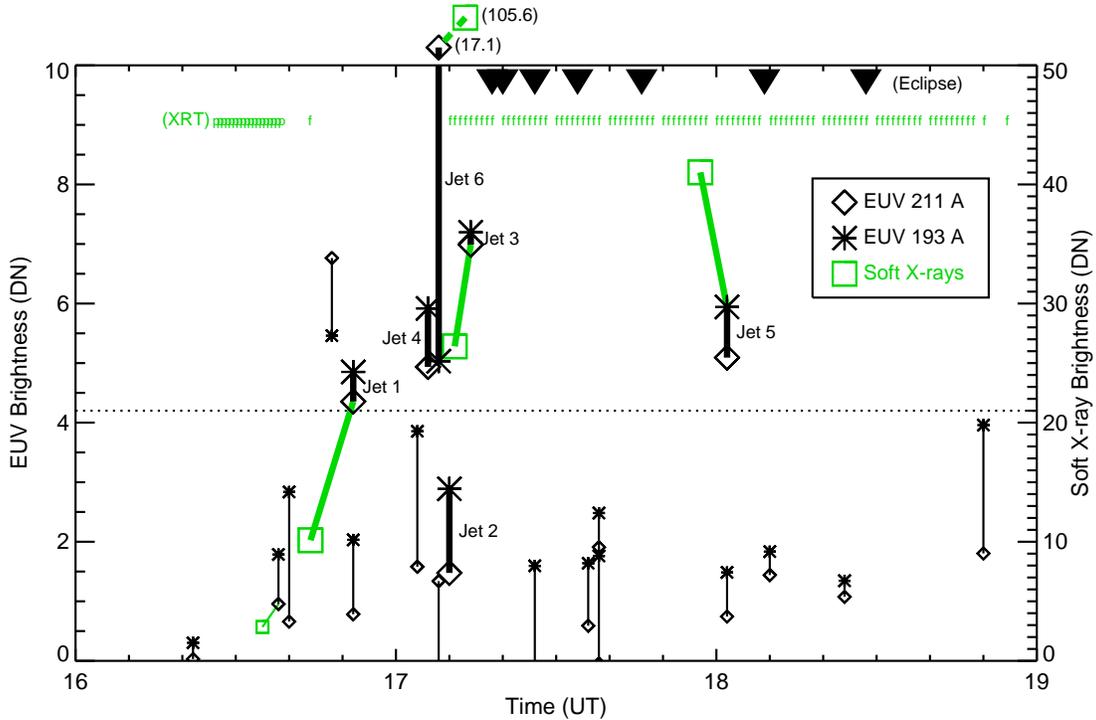}
\caption{Brightness of the EUV and soft X-ray jets plotted at their peak times.  Diamonds, asterisks, and green squares show the brightness values measured on the 211 \AA\ images, 193 \AA\ images, and soft X-ray images.  Symbols of an identical jet are connected by lines.  The jets accompanied by the eclipse jets are displayed with large symbols and thick lines, and jet numbers are added.  The 211 \AA\ and soft X-ray brightness values of jet 6 exceed the plot range, and the values are noted instead.  The eclipse observation epochs are marked with triangles, and the acquisition times of the soft X-ray images are marked with ``p'' for the partial images and with ``f'' for the full-Sun images.  The horizontal dotted line shows the level above which all the EUV jets occurring not too distant from the eclipse observation epochs are accompanied by the eclipse jets.
\label{fig:fig6}}
\end{figure}

As shown in Section 3.1, all the eclipse jets are accompanied by EUV jets.  The question is, what kind of EUV jets develop into high-altitude eclipse jets?  We compared the EUV jets developing into the eclipse jets and those not related to the eclipse jets.  We checked the 211 \AA\ and 193 \AA\ images taken with the AIA around the eclipse time, 16:00 --19:00 UT on August 21, and visually picked EUV jets which clearly extend above the limb in the northern and southern polar regions.  We found 21 jets, and the footpoints of the jets are marked on a 211 \AA\ image in Figure 5(a).  They are mostly located in polar coronal holes.  Red and green arrows correspond to the jets accompanied and not accompanied by the eclipse jets, respectively.  Some of the jets recurred at the same position.

In addition, we checked X-ray jets on the images taken with the XRT of {\it Hinode}, though the coverage of the observation with the XRT was limited temporally and spatially as described in Section 2.  We used the images taken with the ``Al-poly'' filter, because this filter has extended temperature sensitivity, and therefore, the images taken with this filter are often used to search jets; see e.g., \citet{2013ApJ...775...22S}.  We found five jets above the limb in the polar regions, and all of them have the EUV counterparts, with four corresponding to the eclipse jets.  The footpoints of the soft X-ray jets are marked in Figure 5(b).  Again, red and green arrows correspond to the jets accompanied and not accompanied by the eclipse jet, respectively.  

We measured the brightness of the EUV jets on the 211 \AA\ and 193 \AA\ images and that of the soft X-ray jets on the images taken with the Al-poly filter.  The brightness is defined as the data number of the excess brightness of the jets (with respect to the background corona) integrated along the pass vertically crossing the jets at their middle height.  In Figure 6, the brightness of the EUV (black) and soft X-ray (green) jets are plotted at their peak times.  The jets accompanied by the eclipse jets are shown with large symbols and thick lines.  As seen in Figure 6, all of the EUV jets brighter than a certain level (see the dotted line) after 16:50 UT (not too far from the eclipse observation epochs) were found in the eclipse images without exception.  This fact suggests that at least the jets whose brightness exceed a certain level generally reach high altitudes.  

In Figure 6, there is another EUV jet as bright as those accompanied by the eclipse jet, occurring at 16:47 UT, just four minutes before jet 1.  If jets with this level of brightness generally reach high altitudes, it is possible that this jet is recognized in the eclipse images.  This jet is located at position angle $180^\circ$ and it is marked in Figure 5(a) with a large green arrow.  The EUV images of this jet and the corresponding area in the eclipse images (a strip marked with dashed lines in Figure 1) are shown in Figure 3(g).  A darkening can be found in the eclipse images, and it is plausible that this is the declining phase of a high-altitude component of the EUV jet at 16:47 UT.  The occurrence of this jet was probably slightly too early to be recognized as an eclipse jet.  Therefore, this jet also supports the conclusion that at least the jets whose brightness exceed a certain level generally reach high altitudes.

On the other hand, there are many faint EUV jets that were not found in the eclipse images except one (jet 2), as seen in Figure 6.  Did these faint jets reach high altitudes?  We cannot judge that right now, because the faint jets are generally considered to be below the detection limit of the eclipse observation, even if they reach high altitudes.  In our eclipse observation, the time cadence is far sparser than that of the AIA observation. The differences in the instruments and the sky condition of the observing sites limit the signal-to-noise ratio of the difference images.  It is required to carry out the eclipse observation (or another type of observation of the corona) with a higher cadence and a higher signal-to-noise ratio to detect fainter white-light jets.  Among the faint jets, jet 2 is accompanied by an eclipse jet.  As shown in Figure 5(a), small jets recurrently occurred at the position of this jet (almost at the north pole).  Therefore, it is possible that the recurrent supply of hot plasma by the small jets made this jet detectable on the eclipse images.

For the soft X-ray jets, four brighter jets out of five are accompanied by eclipse jets without exception.  Therefore, again we can safely presume that the soft X-ray jets which are at least brighter than a certain level reach high altitudes, and furthermore, reaching high altitudes is possibly a common feature of most of the soft X-ray jets.

\section{Discussion} \label{sec:discussion}

\subsection{Brightness and frequency of the EUV and soft X-ray jets accompanied by the eclipse jets}

White-light polar jets observed with the LASCO C2 and the SECCHI-COR1 are much less frequent \citep{1998ApJ...508..899W, 2009SoPh..259...87N, 2010SoPh..264..365P, 2015ApJ...806...11M} than ordinary polar soft X-ray jets, which can be seen a few times in an hour in the polar regions \citep{2007PASJ...59S.771S, 2013ApJ...775...22S, 2015A&A...579A..96P}.  This means that only some of the jets, probably the energetic ones, have been confirmed to reach high altitudes so far.  On the other hand, our eclipse observations during 70 minutes show that six jets reached high altitudes.  One of the observed jets, jet 6, is possibly as large as those observed with the spaceborne coronagraphs, but the other jets are smaller.  This result implies that even ordinary polar jets can also reach high altitudes.  However, it is not clear if the observed jets correspond to ordinary jets so far observed.  Then, we compared the brightness and frequency of the jets with the results from former statistical analyses.

Firstly, we compared the surface brightness of the soft X-ray jets with the result by \citet{2013ApJ...775...22S}.  The surface brightness measured at the half height of the length of the jets on the soft X-ray images taken with the Al-poly filter is in the range of 0.8 -- 2.0 DNs$^{-1}$pixel$^{-1}$ for the jets accompanied by the eclipse jets and 0.5 DNs$^{-1}$pixel$^{-1}$ for the jet not accompanied by the eclipse jets.  The range of these brightness values corresponds to the brightness where the jets in the polar coronal holes found most frequently \citep{2013ApJ...775...22S}.  Therefore, although the integrated soft X-ray brightness shown in Figure 6 scatters widely, there is no evidence that the observed soft X-ray jets are particularly energetic, except for jet 6, and they are presumed to be within the scope of ordinary soft X-ray jets.

Secondly, we compared the frequency of the jets with the previous studies.  \citet{2013ApJ...775...22S} gave the daily occurrence rate of the soft X-ray jet in the polar coronal holes to be $6.84\times 10^{-12}$ km$^{-2}$ hr$^{-1}$.  Because the footpoints of our target jets are close to the limb, we assumed that the area of the footpoints of the observed jets is limited to $40^\circ$ (along the position angle)$\times 10^\circ$ (depth) $\times 2$ (both poles), and we got the occurrence rate of 0.8 hr$^{-1}$.  This does not contradict the result by \citet{2007PASJ...59S.771S}, 60 soft X-ray jets per day in the whole polar regions, and that by \citet{2015A&A...579A..96P}, 18 soft X-ray jets in nine hours around one of the poles.  On the other hand, the number of eclipse jets found during the 70-minute eclipse observations is six; the number of bright EUV jets found during three hours is six; the number of soft X-ray jets found during three hours with some gaps in the observation is five, and four out of five are accompanied by the eclipse jet.  All of these numbers are consistent with the occurrence rate estimated in the previous studies for the soft X-ray jets.  This fact means that the eclipse jets occur as often as the ordinary soft X-ray jets, and therefore, the ordinary soft X-ray jets are generally presumed to reach high altitudes, while it is still not clear if the faint EUV jets, which occur much more frequently than the soft X-ray jets, reach high altitudes.

Therefore, on the basis of the brightness and the frequency of the observed jets, it is deduced that the eclipse jets, which reach high altitudes and probably eventually escape from the Sun, are not different from ordinary soft X-ray jets in polar regions.  It is possible that a part of the jet material goes back to the solar surface as pointed out by \citet{2007PASJ...59S.751C}, \citet{2007PASJ...59S.771S}, \citet{2014A&A...561A.104C}, and \citet{2015A&A...579A..96P}.  However, the eclipse observations revealed that it is one of the common characteristics of the ordinary polar jets to escape from the Sun as part of the solar wind, even if all the jet material does not necessarily escape.

\subsection{Length, velocity, and lifetime of jets}

The eclipse observations of jets revealed that the ordinary polar jets actually have some characteristics unseen in soft X-ray and EUV images as follows.

The length of the ordinary polar X-ray jets has been measured to be 0.44--16.2$\times 10^4$ km by \citet[][and \citet{2007PASJ...59S.771S} and \citet{2015A&A...579A..96P} gave similar results]{2013ApJ...775...22S}.  The EUV jets accompanied by the eclipse jets shown in Figure 3 have the length of about $10^5$ km.  However, the eclipse jets show that the ordinary polar jets extend beyond the limit of the observations in the soft X-rays and the EUV.

The lifetime of ordinary polar X-ray jets is typically 10 minutes \citep{2013ApJ...775...22S, 2007PASJ...59S.771S}, but it is the characteristic time scale of jets seen in the soft X-rays.  If we take the white-light eclipse jets into consideration, it takes more than 60 minutes from the beginning of the EUV jets to the decay of the eclipse jets, as shown in Figure 3.  This is comparable to 70--80 minutes obtained by \citet{2009SoPh..259...87N}, who measured the lifetime of probably energetic jets observed both with the EUVI and COR1 of SECCHI.

Regarding the outward-propagating velocity of the leading edge of jets, there has been a discrepancy between the soft X-ray jets and the white-light jets observed with spaceborne coronagraphs.  The average velocity of the soft X-ray jet was measured to be 181 km s$^{-1}$ \citep{2013ApJ...775...22S} and 160 km s$^{-1}$ \citep{2007PASJ...59S.771S}.  On the other hand, spaceborne coronagraph observations gave the average or median velocity of 300--600 km s$^{-1}$ \citep{1998ApJ...508..899W, 2009SoPh..259...87N, 2010SoPh..264..365P, 2015ApJ...806...11M}.  The high velocity of the white-light jets might be explained as being estimated from the particularly energetic samples.  However, in our study shown in Table 2, the average velocity of the eclipse jets seen in the EUV images ($v_\mathrm{EUV}$ in Table 2) is 360 km s$^{-1}$, and that estimated from the length of the eclipse jets and the time elapsed from the occurrence of the EUV jets ($v_{\mathrm{EUV}+\mathrm{eclipse}}$ in Table 2) is 450 km s$^{-1}$.  This means that the different velocities in the EUV and the white light appear in the same jets.  It is a plausible interpretation of this fact that the jets are accelerated during their upward movement.  However, it should also be noted that the velocity measured using only the images of the eclipse ($v_\mathrm{eclipse}$ in Table 2) tends to show smaller values than $v_{\mathrm{EUV}+\mathrm{eclipse}}$.  Defining the leading edge of jets is strongly affected by the visibility of jets, and the faint leading edge of jets can be missed.  \citet{2007PASJ...59S.771S} reported that a kind of velocity estimation methods gave the velocity of 600--1000 km s$^{-1}$ for soft X-ray jets, while other methods gave the velocity of 70--400 km s$^{-1}$.  Therefore, we need to pay attention to the strong dependence of the estimated velocity on the observation method and the analysis method.

Although various parameters obtained based on the soft X-ray observations are still appropriate as characteristic values for the soft X-ray jets, it should be noted that the ultimate height, lifetime and true leading-edge velocity of jets, which are difficult to be obtained only from the soft X-ray observations, might be required to consider the mechanism of jet production.

\subsection{Relation to the solar wind}

We revealed that the ordinary polar jets generally reach high altitudes and escape from the Sun.  This means that jets fainter than those so far considered to escape from the Sun contribute to the solar wind.  To estimate mass flux caused by the ordinary jets is important in considering the contribution of jets to the solar wind.  As seen in Figure 4, the eclipse jets show brightness changes of the order of one percent with respect to the background corona around the height of 1.5--2 $R_\odot$, while those at 2.5 $R_\odot$ are smaller.   The brightness of the background corona comes from both the K- and F-corona.  The brightness changes with respect to the background K-corona are estimated to be 2.5 percent at 1.5 $R_\odot$ and five percent at 2 $R_\odot$ on the assumption that the fraction of the K-corona is 0.4 at 1.5 $R_\odot$ and 0.2 at 2 $R_\odot$ \citep{2000asqu.book.....C}.  The brightness of the background K+F corona is 3.5--1.0$\times 10^{-8}$ $B_\odot$ at 1.5--2 $R_\odot$ in our measurement, and the excess brightness due to jets is inferred to be typically 3.5--1.0$\times 10^{-10}$ $B_\odot$, where $B_\odot$ is the average brightness of the white-light solar disk.  Because the typical width of the jets is measured to be 2.2$\times 10^9$ cm on the eclipse difference images, the electron density is estimated to be 1.5--1.0$\times 10^6$ cm$^{-3}$ based on the Thomson scattering efficiency.  From the above values, electron flux passing through the section of the jet at 1.5--2 $R_\odot$ is estimated to be 1.8--1.2$\times 10^{32}$ s$^{-1}$ on the assumption of the bulk velocity to be 225 km s$^{-1}$.  The velocity of 225 km s$^{-1}$ is assumed to be half of the typical velocity of the leading edge, according to the estimation by \citet{1998ApJ...508..899W}.  \citet{2013ApJ...775...22S} estimated the thermal energy of a polar coronal hole jet to be 9$\times 10^{25}$ erg, and this gives the total electron number of 2.2$\times 10^{35}$ on the assumption that the temperature of jets is $10^6$ K.  If the eclipse jets with peak brightness survive for 1300--2000 s, the electron flux reaches this value.  This does not contradict to the measured total duration of eclipse jet of more than 60 minutes, because the average brightness of the jets during the lifetime is lower than the peak brightness.  Therefore, the mass flux of each jet estimated on the basis of the eclipse observation is consistent with the previous estimation based on the soft X-ray observations. 

As described above, in the EUV images, we found many faint jets, which were not found in the eclipse images.  Furthermore, \citet{2008ApJ...682L.137R} and \citet{2014ApJ...787..118R} pointed out that numerous smaller jetlike phenomena (referred as jetlets) can be found in EUV images taken with the AIA, and we can also find many small spoutings from bright points inside the limb in the polar region in the EUV images during the eclipse period, in addition to the jets above the limb.  Therefore, how small jets and jetlets contribute to the solar wind is a subject to be studied in future.  However, it should be noted that \citet{2013ApJ...775...22S} pointed out that the integrated energy of jets contributes to the solar wind for only about one percent, even if the energy of jet is integrated down to the unseen smallest jets on the assumption that all the jets contribute to the solar wind.  The estimation based on  the observations of jets by \citet{2014ApJ...784..166Y} and \citet{2013ApJ...776...16P} gave the results similar to that of \citet{2013ApJ...775...22S}, and based on a magnetohydrodynamic simulation, \citet{2016ApJ...831L...2L} also obtained the result that the contribution of jets to the mass and energy of the solar wind is around one percent.  As shown above, our analysis shows that the mass flux of the eclipse jets is consistent with that of the soft X-ray jets.  Therefore, our results support the conclusion that the polar jets are not major contributors to the fast solar wind originating from the polar coronal holes.

\subsection*{}

Summarizing the above, on the basis of a time-series white-light images taken at the solar eclipse and EUV, soft X-ray, and white-light data taken with spaceborne instruments, we found that ordinary jets occurring in the polar coronal holes actually extend to high altitudes and presumably eventually escape from the Sun as part of fast solar wind.  Such results are brought about by the combination of the observation of the solar eclipse and that with the spaceborne instruments, which enabled us to fill the observation gap of the spaceborne instruments around 1.2--2.0 $R_\odot$.  To observe the corona in this height range is still more promising to gain further understanding of the corona.

\acknowledgments

The AIA data used here are provided courtesy of NASA/SDO and the AIA science team. {\it Hinode} is a Japanese mission developed and launched by ISAS/JAXA, with NAOJ as domestic partner and NASA and STFC (UK) as international partners. It is operated by these agencies in co-operation with ESA and NSC (Norway).  The SOHO/LASCO data used here are produced by a consortium of the Naval Research Laboratory (USA), Max-Planck-Institut f\"ur Aeronomie (Germany), Laboratoire d'Astronomie Spatiale (France), and the University of Birmingham (UK). SOHO is a project of international cooperation between ESA and NASA.


\begin{thebibliography}{}

\bibitem[Brueckner et al.(1995)]{1995SoPh..162..357B} Brueckner, G.~E., Howard, R.~A., Koomen, M.~J., et al.\ 1995, \solphys, 162, 357 

\bibitem[Chandrashekhar et al.(2014)]{2014A&A...561A.104C} Chandrashekhar, K., Bemporad, A., Banerjee, D., Gupta, G.~R., \& Teriaca, L.\ 2014, \aap, 561, A104 

\bibitem[Cox(2000)]{2000asqu.book.....C} Cox, A.~N.\ 2000, Allen's Astrophysical Quantities, ed. A.~N. Cox (New York: Springer) 

\bibitem[Culhane et al.(2007)]{2007PASJ...59S.751C} Culhane, L., Harra, L.~K., Baker, D., et al.\ 2007, \pasj, 59, S751 

\bibitem[Delaboudini{\`e}re et al.(1995)]{1995SoPh..162..291D} Delaboudini{\`e}re, J.-P., Artzner, G.~E., Brunaud, J., et al.\ 1995, \solphys, 162, 291 

\bibitem[Domingo et al.(1995)]{1995SoPh..162....1D} Domingo, V., Fleck, B., \& Poland, A.~I.\ 1995, \solphys, 162, 1 

\bibitem[Golub et al.(2007)]{2007SoPh..243...63G} Golub, L., Deluca, E., Austin, G., et al.\ 2007, \solphys, 243, 63 

\bibitem[Hanaoka et al.(2012)]{2012SoPh..279...75H} Hanaoka, Y., Kikuta, Y., Nakazawa, J., Ohnishi, K., \& Shiota, K.\ 2012, \solphys, 279, 75 

\bibitem[Hanaoka et al.(2014)]{2014SoPh..289.2587H} Hanaoka, Y., Nakazawa, J., Ohgoe, O., Sakai, Y., \& Shiota, K.\ 2014, \solphys, 289, 2587 

\bibitem[Howard et al.(2008)]{2008SSRv..136...67H} Howard, R.~A., Moses, J.~D., Vourlidas, A., et al.\ 2008, \ssr, 136, 67 

\bibitem[Kaiser et al.(2008)]{2008SSRv..136....5K} Kaiser, M.~L., Kucera, T.~A., Davila, J.~M., et al.\ 2008, \ssr, 136, 5 

\bibitem[Kosugi et al.(2007)]{2007SoPh..243....3K} Kosugi, T., Matsuzaki, K., Sakao, T., et al.\ 2007, \solphys, 243, 3 

\bibitem[Lemen et al.(2012)]{2012SoPh..275...17L} Lemen, J.~R., Title, A.~M., Akin, D.~J., et al.\ 2012, \solphys, 275, 17 

\bibitem[Lionello et al.(2016)]{2016ApJ...831L...2L} Lionello, R., T{\"o}r{\"o}k, T., Titov, V.~S., et al.\ 2016, \apjl, 831, L2 

\bibitem[Moore et al.(2010)]{2010ApJ...720..757M} Moore, R.~L., Cirtain, J.~W., Sterling, A.~C., \& Falconer, D.~A.\ 2010, \apj, 720, 757 

\bibitem[Moore et al.(2015)]{2015ApJ...806...11M} Moore, R.~L., Sterling, A.~C., \& Falconer, D.~A.\ 2015, \apj, 806, 11 

\bibitem[Neugebauer(2012)]{2012ApJ...750...50N} Neugebauer, M.\ 2012, \apj, 750, 50 

\bibitem[Nistic{\`o} et al.(2009)]{2009SoPh..259...87N} Nistic{\`o}, G., Bothmer, V., Patsourakos, S., \& Zimbardo, G.\ 2009, \solphys, 259, 87 

\bibitem[Paraschiv et al.(2010)]{2010SoPh..264..365P} Paraschiv, A.~R., Lacatus, D.~A., Badescu, T., et al.\ 2010, \solphys, 264, 365 

\bibitem[Paraschiv et al.(2015)]{2015A&A...579A..96P} Paraschiv, A.~R., Bemporad, A., \& Sterling, A.~C.\ 2015, \aap, 579, A96 

\bibitem[Pasachoff et al.(2008)]{2008ApJ...682..638P} Pasachoff, J.~M., Ru{\v s}in, V., Druckm{\"u}ller, M., et al.\ 2008, \apj, 682, 638 

\bibitem[Pesnell et al.(2012)]{2012SoPh..275....3P} Pesnell, W.~D., Thompson, B.~J., \& Chamberlin, P.~C.\ 2012, \solphys, 275, 3 

\bibitem[Pucci et al.(2013)]{2013ApJ...776...16P} Pucci, S., Poletto, G., Sterling, A.~C., \& Romoli, M.\ 2013, \apj, 776, 16 

\bibitem[Raouafi et al.(2008)]{2008ApJ...682L.137R} Raouafi, N.-E., Petrie, G.~J.~D., Norton, A.~A., Henney, C.~J., \& Solanki, S.~K.\ 2008, \apjl, 682, L137 

\bibitem[Raouafi \& Stenborg(2014)]{2014ApJ...787..118R} Raouafi, N.-E., \& Stenborg, G.\ 2014, \apj, 787, 118 

\bibitem[Raouafi et al.(2016)]{2016SSRv..201....1R} Raouafi, N.~E., Patsourakos, S., Pariat, E., et al.\ 2016, \ssr, 201, 1 

\bibitem[Sako et al.(2013)]{2013ApJ...775...22S} Sako, N., Shimojo, M., Watanabe, T., \& Sekii, T.\ 2013, \apj, 775, 22 

\bibitem[Savcheva et al.(2007)]{2007PASJ...59S.771S} Savcheva, A., Cirtain, J., Deluca, E.~E., et al.\ 2007, \pasj, 59, S771 

\bibitem[Shibata et al.(1992)]{1992PASJ...44L.173S} Shibata, K., Ishido, Y., Acton, L.~W., et al.\ 1992, \pasj, 44, L173 

\bibitem[Shimojo et al.(1996)]{1996PASJ...48..123S} Shimojo, M., Hashimoto, S., Shibata, K., et al.\ 1996, \pasj, 48, 123 

\bibitem[Wang et al.(1998)]{1998ApJ...508..899W} Wang, Y.-M., Sheeley, N.~R., Jr., Socker, D.~G., et al.\ 1998, \apj, 508, 899 

\bibitem[Wood et al.(1999)]{1999ApJ...523..444W} Wood, B.~E., Karovska, M., Cook, J.~W., Howard, R.~A., \& Brueckner, G.~E.\ 1999, \apj, 523, 444 

\bibitem[Wuelser et al.(2004)]{2004SPIE.5171..111W} Wuelser, J.-P., Lemen, J.~R., Tarbell, T.~D., et al.\ 2004, \procspie, 5171, 111 

\bibitem[Yu et al.(2014)]{2014ApJ...784..166Y} Yu, H.-S., Jackson, B.~V., Buffington, A., et al.\ 2014, \apj, 784, 166 


\end{thebibliography}
\end{document}